\documentclass[twocolumn,prX]{revtex4-1}

\usepackage{graphicx}
\usepackage{graphics}
\usepackage{epstopdf}

\begin{document}

 \title{Energy transfer among distant quantum systems
       in spatially shaped laser fields: \\
       Two H atoms 
       with the internuclear separation 
       of 5.29~nm (100~a.u.)}

 \author{{Guennaddi K.~Paramonov, Oliver K\"uhn \\}
 {\em Institut f\"ur Physik, Universit\"at Rostock,
  18051 Rostock, Germany} \\
  and Andr\'e D.~Bandrauk \\
  {\em Laboratorie de Chimie Th\'eorique, Facult\'e des Sciences, 
   Universit\'e de Sherbrooke, Sherbrooke, Qu\'ebec, Canada J1K 2R1}}
      
 \begin{abstract}
 The quantum dynamics of two distant H atoms excited
 by ultrashort and spatially shaped laser pulses 
 is studied by the numerical solution of 
 the non-Born-Oppenheimer time-dependent Schr\"odinger equation
 within a three-dimensional (3D) model,
 including the internuclear distance $R$ and
 the two $z$ coordinates of the electrons, $z_{1}$ and $z_{2}$. 
 The two 1D hydrogen atoms, {\rm A} and {\rm B}, are assumed to be initially
 in their ground states 
 with a large (but otherwise arbitrary) internuclear separation of $R=100$~a.u. (5.29~nm).
 Two types of a spatial envelope of a laser field linearly polarized 
 along the $z$-axis are considered:
 (i) a broad Gaussian envelope, such that atom~{\rm A} is excited 
 by the laser field predominantly, 
 and (ii) a narrow envelope, such that practically only atom~{\rm A} is excited
 by the laser field.
 With the laser carrier frequency $\omega =1.0$~a.u. 
 and the pulse duration $t_{p}=5$~fs,
 in both cases an efficient energy transfer 
 from atom~{\rm A} to atom~{\rm B} has been found.
 The ionization of atom~{\rm B}  
 achieved mostly after the end of the laser pulse 
 is close to or even higher than that of atom~{\rm A}.
 It is shown that with a narrow spatial envelope of the laser field, 
 the underlying mechanisms of the 
 energy transfer from {\rm A} to {\rm B} and the ionization of {\rm B}
 are the Coulomb attraction of the laser driven electron
 by the proton of atom~{\rm B} 
 and a short-range Coulomb repulsion of the two electrons
 when their wave functions significantly overlap in the domain of atom~{\rm B}.
 In the case of a broad Gaussian spatial envelope of the laser field,  
 the opposite process also occurs, but with smaller probability:
 the energy is transferred from the weakly excited atom~{\rm B} to atom~{\rm A},
 and the ionization of atom~{\rm A} is also induced by the electron-electron repulsion 
 in the domain of atom~{\rm A} due to a strong overlap of the electronic wave functions.
 
 \end{abstract}

 \maketitle

 \section{Introduction} 

 The laser driven dynamics 
 of distant quantum systems
 can depend in general on
 the spatial and temporal envelope of the applied laser field.
 For example, at a gas pressure of 1~atm., the interparticle distance
 is about 100~a.u. (5.29~nm). If such a system, e.g. composed of H atoms, 
 is excited by a laser field with the carrier frequency $\omega =1.0$~a.u.,
 corresponding to the ground-state energy of H$_{2}$ 
 at a large internuclear distance $R$, 
 the wavelength is $\lambda =861$~a.u. (45.56~nm).
 If the laser field is focused within the diffraction limit
 onto a spot with the width $\lambda $,
 the Gaussian spatial envelope of the field 
 may result in quite
 different electric field strengths for H atoms
 separated by about 100~a.u.,
 especially at the edges of the Gaussian spatial envelope. 
 Although the H atoms are far away from each other, their electron-electron
 interaction should not be {\em a~priori} neglected, 
 especially upon their excitation by the laser field,
 because the electronic wave functions extend and vanish, strictly speaking,
 only at infinity. Therefore, the energy transfer among distant quantum systems,
 similar to that studied in
 \cite{Cederbaum:97prl.ICD,%
 Marburger:03prl.ICD,%
 Jahnke:04prl.ICD,%
 Morishita:06prl.ICD,%
 Lablanquie:07jcp.ICD,%
 Jahnke:10NaturePhys.ICD,%
 Havermeier:10prl.ICD,%
 Sisourat:10NatPhys.ICD},
 can be anticipated to occur in spatially shaped laser fields as well.
 For ultrashort laser pulses, containing only few optical cycles,
 one must also consider the carrier envelope phase (CEP) of the pulse
 \cite{Bandrauk:2002.PRA.HatomNc15}.

 The long-range energy transfer from an excited atom to its neighbor has been recently studied
 by Cederbaum {\em et al.} for molecular clusters
 \cite{Cederbaum:97prl.ICD}
 and is known as the interatomic Coulombic decay (ICD).
 Nowadays, ICD is well established also experimentally 
 for inner-valence excitation
 of many electron systems
 \cite{Marburger:03prl.ICD,%
 Jahnke:04prl.ICD,%
 Morishita:06prl.ICD,%
 Lablanquie:07jcp.ICD,%
 Jahnke:10NaturePhys.ICD}.
 In recent work
 \cite{Havermeier:10prl.ICD},
 ICD was demonstrated experimentally for a helium dimer.
 Since helium atoms have no inner-valence electrons, 
 a different type of ICD is operative for this case.
 It was thus concluded in
 \cite{Havermeier:10prl.ICD}
 that since ICD in a helium dimer takes place at interatomic distances
 up to $\approx 12$~a.u., no overlap of the electronic wave functions
 is required for the process.

 The present work is addressed to a quantum system composed of two H atoms
 with the initial internuclear separation of 100~a.u. (5.29~nm)
 which is excited by spatially shaped laser pulses: spatially broad pulses
 exciting both H atoms, and spatially narrow pulses exciting
 only one H atom of the entire H-H system.
 The relative simplicity of the H-H system under consideration
 (similar to that used in 
 \cite{Bandr:08prl.H2d3})
 makes it possible to treat
 the long-range electronic motion explicitly together with the nuclear motion
 such as to reveal the role played by the electron-electron interaction
 and by the overlap of the electronic wave functions.
 An example of long-range laser-induced electron transfer (LIET) 
 in the one electron linear H$^{+}$-H$_{2}^{+}$ atom-molecule system 
 has been treated previously
 within the Born-Oppenheimer approximation
 \cite{Bandr:07prl.HpH2pBO}.
 Long-range charge and energy transfer can occur also in large molecular systems,
 as described recently in Ref.
 \cite{MayKuehn:2010book} 
 and references therein.

 The following two types of H-H systems will be distinguished in the present work:
 (i) a `molecular' H-H system,
 representing an elongated
 configuration of the H$_{2}$ molecule, similar to that studied recently in 
 \cite{BandrShon:pra.VPA.Entang}
 for long-range entanglement,
 and (ii) an `atomic' H-H system, representing two distant H atoms.
 Accordingly, the initial state of a molecular H-H system  
 is assumed to be entangled by spin exchange and represented by the Heitler-London
 symmetric product of atomic wave functions,
 while the initial state of an atomic H-H system is not entangled --
 it is a direct-product state of atomic wave functions. 
 In both cases the excitation of H-H is accomplished
 by laser pulses with (i) a broad Gaussian spatial envelope,
 such that both H atoms are excited by the laser field, 
 with atom~{\rm A} being excited predominantly,
 and (ii) with a narrow spatial envelope,
 such that only atom~{\rm A} is excited by the laser field.

 The paper is organized as follows. The model of the H-H system and techniques used
 are described in Sec.~II. 
 Excitation, energy transfer, and ionization of an unentangled atomic H-H system 
 are presented in Sec.~III. Section~IV is devoted
 to the laser-driven dynamics of an entangled molecular H-H system. 
 The results obtained are summarized and discussed in the concluding Section~V.

 \section{Model, equations of motion, and techniques} 

 Within the 3D four-body model of H-H
 excited by the temporally and spatially shaped laser field
 the total Hamiltonian $\hat H_{\rm T}$ 
 is divided into two parts,
 \begin{equation}
 \hat H_{\rm T}(R,z_{1},z_{2},t) = \hat H_{\rm S}(R,z_{1},z_{2}) + \hat H_{\rm SF}(z_{1},z_{2},t),
  \label{E-2}
 \end{equation}
 where $\hat H_{\rm S}(R,z_{1},z_{2})$ represents the H-H system and
 $\hat H_{\rm SF}(z_{1},z_{2},t)$ describes the interaction of the system
 with the laser field. 
 The applied laser field is assumed to be linearly polarized along
 the $z$-axis, the nuclear and the electronic motion are restricted
 to the polarization direction of the laser electric field.
 Accordingly, two $z$ coordinates of electrons, $z_{1}$ and $z_{2}$,
 measured with respect to the nuclear center of mass,
 are treated explicitly together with the internuclear distance $R$. 
 A similar model has been used previously in
 \cite{Bandr:08prl.H2d3}
 for the H$_{2}$ molecule, where each particle, electron or proton,
 is treated in 1D, i.e., $z$ and $R$.

 The total non-Born-Oppenheimer system Hamiltonian 
 (employing a. u.: $e=\hbar=m_{e}=1$) reads
 \begin{displaymath}
 \hat H_{\rm S}(R,z_{1},z_{2}) =
 -\frac{1}{m_{\rm p}}\frac{\partial^{2}}{\partial R^{2}} 
 + V_{\rm p p}(R) 
 \end{displaymath}
 \begin{equation}
 + \sum _{k=1}^{2} \biggl[-\frac{1}{2 \mu_{\rm e}}\frac{\partial^{2}}{\partial z_{k}^{2}} 
  + V_{\rm e p}(z_{k},R)\biggr] + V_{\rm e e}(z_{1},z_{2}),
  \label{E-3}
 \end{equation}
 where $m_{\rm p}$ is the proton mass, $\mu_{\rm e}=2m_{\rm p}/(2m_{\rm p}+1)$
 is the reduced electron mass, and non-diagonal mass-polarization terms
 are neglected.
 The Coulomb potentials in Eq.~(\ref{E-3}) read
 \begin{displaymath}
 V_{\rm p p}(R)=\frac{1}{R}, \, \, \, \, \, \, \, 
 V_{\rm e e}(z_{1},z_{2})=\frac{1}{\sqrt{(z_{1}-z_{2})^{2}+\alpha}},
 \end{displaymath}
 \begin{equation}
 V_{\rm e p}(z_{k},R)=-\frac{1}{\sqrt{(z_{k}-R/2)^{2}+\beta}}-\frac{1}{\sqrt{(z_{k}+R/2)^{2}+\beta}},
  \label{E-4}
 \end{equation}
 where $k=1,2$, and the regularization parameters,  $\alpha=0.1\times 10^{-3}$ and $ \beta=1.995$, 
 have been chosen (similar to previous work
 \cite{Bandr:08prl.H2d3})
 such as to reproduce the ground-state (GS) energy  
 of the H-H system at $R=100$~a.u.
 (${\rm E}_{\rm H-H}^{\rm GS}=-1.0$~a.u.).

 The interaction of the H-H system with the laser field  is treated 
 within the semiclassical electric dipole approximation by the Hamiltonian
 \begin{equation}
 \hat H_{\rm SF}(z_{1},z_{2},t) = -\frac{1}{c}\frac{\partial A(t)}{\partial t}
  (1+\gamma)
   \sum _{k=1}^{2} F(z_{k}) z_{k}, 
  \label{E-6}
 \end{equation}
 where $\gamma=(1+2m_{\rm p})^{-1}$,  
 $A(t)$ is the vector potential, $c$ is the speed of light,
 and $F(z)$ is the spatial-shape function, or envelope, of the laser field.
 In the general case we set $F(z_{1})=F(z_{2})$,
 except for some model simulations detailed in Sec.~IV.

 The vector potential, $A(t)$,  is chosen in the following form:
 \begin{equation}
 A(t)=\frac{c}{\omega}{\cal E}_{0}\sin^{2}(\pi t/t_{p})\cos(\omega t + \phi),
  \label{E-7}
 \end{equation}
 where ${\cal E}_{0}$ is the amplitude, 
 $t_{p}$ is the pulse duration at the base,
 $\omega $ is the laser carrier frequency,
 and $\phi$ is the carrier-envelope phase (CEP). 
 Note that it has been shown previously
 \cite{Bandrauk:2002.PRA.HatomNc15}
 that the carrier phase of the laser pulse is important only for pulses
 having less than 15 optical cycles.
 The definition of the system-field interaction by Eq.~(\ref{E-6}) via the vector potential, 
 suggested in  \cite{BandrShon:pra.VPA.Entang},
 assures that the electric field
 ${\cal E}(t)=-\frac{1}{c}\partial A(t)/\partial t$
 has a vanishing direct-current 
 component, $\int_{0}^{t_{p}}{\cal E}(t)dt=0$,
 and satisfies Maxwell's equations in the propagation region.

 It is suitable to define, on the basis of  Eqs.~(\ref{E-6}) and (\ref{E-7}),
 the local effective-field amplitudes for atoms {\rm A} and {\rm B}
 as follows:
 \begin{equation}
 {\cal E}_{0}^{\rm{\rm A}}={\cal E}_{0} F(z_{\rm A}), \; \; \;
 {\cal E}_{0}^{\rm{\rm B}}={\cal E}_{0} F(z_{\rm B }), 
  \label{E-Eampl}
 \end{equation}
 where  $z_{\rm A}=-50$~a.u., $z_{\rm B}=50$~a.u..
 The respective time-dependent electric fields acting on atoms {\rm A} and {\rm B} read
 \begin{displaymath}
 {\cal E}^{{\rm A,B}}(t)={\cal E}_{0}^{{\rm A,B}}[\sin^{2}(\pi t/t_{p})\sin(\omega t + \phi)
 \end{displaymath}
 \begin{equation}
 -\frac{\pi}{\omega t_{p}}\sin(2\pi t/t_{p})\cos(\omega t + \phi)].
  \label{E-Etime}
 \end{equation}
 The first term in Eq.~(\ref{E-Etime}) corresponds to a laser pulse
 with a sin$^{2}$-type temporal envelope,
 while the second, the so-called `switching' term, appears
 due to the finite pulse duration
 \cite{BandrShon:pra.VPA.Entang,%
 Doslic:06pra.VPA}. 

 The 3D time-dependent Schr\"odinger equation for 1D electrons ($z_{1}$, $z_{2}$)
 and 1D protons ($R$),
 \begin{equation}
 i\frac{\partial}{\partial t} \Psi = 
 \bigl[\hat H_{\rm S}(R,z_{1},z_{2})+\hat H_{\rm SF}(z_{1},z_{2},t)\bigr] \Psi ,
  \label{TDSE}
 \end{equation}
 has been solved numerically with the propagation technique adapted from
 \cite{Paramon:05cpl.HH-HD.07cp.HH-HD-Muon}
 for both electron and proton quantum motion.
 In particular, calculations for the electron motion have been performed
 by using 200-point non-equidistant grids
 for the Hermite polynomials and corresponding weights for the numerical
 integration on the interval $(-\infty, \infty)$ for the $z_{1}$ and $z_{2}$
 coordinates. For the nuclear coordinate $R$, a 256-point equidistant grid
 has been used on the interval $[75$~a.u., $125$~a.u.$]$. 
 The time-step of the propagation was $\Delta t = 0.021$~a.u. (1~a.u.=24~asec).
 
 The wave functions of the initial states have been obtained by numerical
 propagation of the equation of motion (\ref{TDSE}) in imaginary time 
 without the laser field (${\cal E}_{0}=0$).

 Upon excitation of the H-H system by the laser field, 
 the electronic wave functions of 
 its atomic {\rm A} and {\rm B} parts may overlap.
 In order to study the energy transfer from {\rm A} to {\rm B},
 the respective `atomic' energies, $E_{\rm A}(t)$ and $E_{\rm B}(t)$, 
 are defined  
 on the basis of Eqs.~(\ref{E-3}) and (\ref{E-4}) such that their sum
 always gives the correct total energy of the entire H-H system.
 The definitions of `atomic' energies $E_{\rm A}(t)$ and $E_{\rm B}(t)$
 for the H-H systems of atomic and molecular origin are different 
 due to the different symmetry and entanglement of the respective wave functions
 and will be specified in the following sections.

 The ionization probabilities for atoms {\rm A} and {\rm B} 
 have been calculated 
 from the time- and space-integrated outgoing fluxes 
 separately for the positive and the negative directions
 of the $z_{1}$ and $z_{2}$ axes 
 at $z_{1,2}=\pm 91$~a.u..
 Specifically, we calculated four ionization probabilities:
 ${\rm I}_{\rm A}(z_{1}=-91\,{\rm a.u.})$ and ${\rm I}_{\rm A}(z_{2}=-91\,{\rm a.u.})$
 for atom {\rm A}, and
 ${\rm I}_{\rm B}(z_{1}=91\,{\rm a.u.})$ and ${\rm I}_{\rm B}(z_{2}=91\,{\rm a.u.})$
 for atom {\rm B}.
 At the outer limits of the $z$-grids, absorbing boundaries have been
 provided by imaginary smooth optical potentials adapted from 
 that designed in
 \cite{Rabitz:94.jcp.HF}. 
 Similar optical potentials have been also provided for the $R$-axis
 but, in practice, the wave-packet never approached the outer limits
 of the $R$-grid.

 \section{Excitation of H-H from an unentangled direct-product initial state} 

 The spatial part of the initial unentangled direct-product ground-state wave function of H-H 
 of an atomic configuration 
 used in the imaginary time propagation is defined with the unsymmetrized 
 Heitler-London electron wave function as follows:
 \begin{equation}
 \Psi(R,z_{1},z_{2},t=0)=\Psi_{1S_{\rm A}}(z_{1}) \Psi_{1S_{\rm B}}(z_{2}) \Psi_{\rm G}(R),
  \label{ini-DP}
 \end{equation}
 where $\Psi_{1S_{\rm A}}(z_{1})=e^{-|z_{1}-z_{\rm A}|}$  at  $z_{\rm A}=-50$~a.u.,
 $\Psi_{1S_{\rm B}}(z_{2})=e^{-|z_{2}-z_{\rm B}|}$ at $z_{\rm B}=50$~a.u.,
 and $\Psi_{\rm G}(R)$ is a proton Gaussian function centered at $R=100$~a.u..

 The imaginary-time propagations have been performed 
 with a reduced, `non-interacting' atoms, version 
 of the system Hamiltonian (\ref{E-3}) wherein 
 the nuclear $V_{\rm p p}(R)$ and electronic $V_{\rm e e}(z_{1},z_{2})$ terms were omitted 
 and only the Coulombic interaction $V_{\rm e p}(z_{1(2)},R)$
 of each electron with its nearest proton was taken into account
 along with the three kinetic-energy terms of Eq.~(\ref{E-3}). 
 The results obtained gave a direct-product initial state.
 For the sake of comparison, we have also performed the imaginary-time propagations
 with the complete system Hamiltonian (\ref{E-3}).
 The results obtained proved to be practically identical to those 
 obtained in the case of two non-interacting H atoms.
 The energy difference, for example, is less than $1.8\times 10^{-5}$~a.u.,
 implying that the Coulombic interaction $V_{\rm e e}(z_{1},z_{2})$ of two H atoms 
 in their ground states is negligible at $R=100$~a.u..

 \begin{figure}[h]
 \includegraphics*[width=16pc]{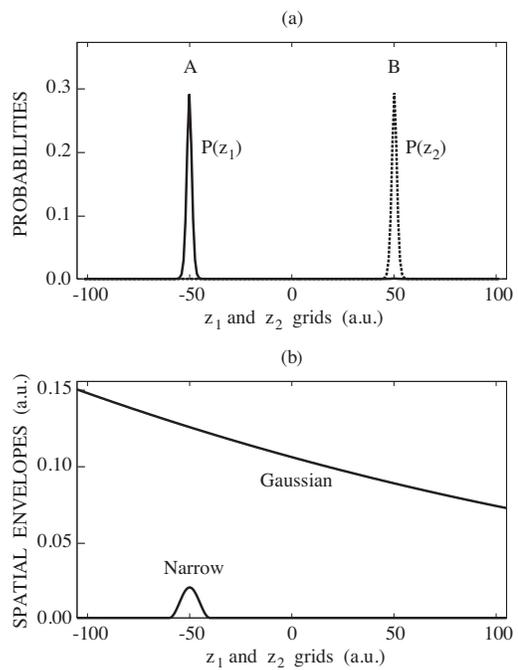}
 \caption{The initial unentangled direct-product state 
 of the H-H system and the spatial envelopes of the applied laser pulses.
 (a) - electron probabilities: 
 $P(z_{1})$ is the probability to find electron $e_{1}$ initially belonging 
 to atom {\rm A} at $z=z_{1}$ 
 with any values of the other two coordinates, $R$ and $z_{2}$; 
 probability $P(z_{2})$ has a similar meaning for electron $e_{2}$ 
 initially belonging to atom {\rm B};
 (b) - spatial envelopes of applied laser fields: 
 the broad Gaussian envelope of Eq.~(\ref{E-G}) 
 and the narrow envelope of Eq.~(\ref{E-N}).}
 \end{figure}

 The initial unentangled direct-product state of the H-H system representing
 two non-interacting H atoms, {\rm A} and {\rm B}, is presented in Fig.~1(a) 
 by the electron probabilities $P(z_{1})$ and $P(z_{2})$, which are defined
 as follows:
 \begin{equation}
 P(z_{1})=\int dR \int dz_{2} |\Psi (R,z_{1},z_{2})|^{2} 
  \label{E-5A}
 \end{equation}
 for electron $e_{1}$ initially belonging to atom {\rm A} with proton $p_{\rm A}$, 
 and similarly,
 \begin{equation}
 P(z_{2})=\int dR \int dz_{1} |\Psi (R,z_{1},z_{2})|^{2},
  \label{E-5B}
 \end{equation}
 for electron $e_{2}$ 
 initially belonging to atom {\rm B} with proton $p_{\rm B}$.
 These electron probabilities give the overall probability to find an electron
 at a specified point of the $z$-axis at any position of the other electron 
 and at any internuclear distance. 

 After calculation of the unentangled initial state
 of the atomic H-H system, its laser-driven quantum dynamics in real time has
 been explored with the complete system Hamiltonian of Eq.~(\ref{E-3}).
 The spatial envelopes of the applied laser pulses, Gaussian and narrow,
 defined by Eqs.~(\ref{E-G}) and Eq.~(\ref{E-N}) below,
 are presented in Fig.~1(b) to illustrate the local effective-field
 strengths acting on atoms {\rm A} and {\rm B}.

 Two possible choices for the laser carrier frequency are:
 $\omega =0.5$~a.u. (corresponding to the ground-state energy of an H atom),
 and $\omega =1.0$~a.u. (corresponding to the ground-state energy of H-H at large $R$).
 Our numerical simulations have shown that at $\omega =0.5$~a.u.,
 an efficient ionization of {\rm A} takes place, 
 while the energy
 transfer from {\rm A} to {\rm B} and the ionization of {\rm B} are not efficient.
 In contrast, at $\omega =1.0$~a.u., the energy transfer from {\rm A} to {\rm B} 
 and the ionization of {\rm B} are efficient. 
 This implies the existence of an optimal laser frequency suitable
 for the most efficient energy transfer. 
 This issue will be addressed in a forthcoming work; 
 below we present the results obtained at $\omega =1.0$~a.u..
 With the laser pulse duration $t_{p}=5$~fs,
 as used throughout this work, the number of optical cycles during the pulse,
 $N_{c}=\omega t_{p}/2\pi$, is about 33. Therefore, the carrier phase of
 the laser field $\phi$ is not important 
 \cite{Bandrauk:2002.PRA.HatomNc15}  
 and set equal to zero in our simulations below [$\phi=0$ in Eqs.~(\ref{E-7}) and (\ref{E-Etime})].

 \begin{figure*}[t]
 \begin{center}
 \includegraphics*[width=30pc]{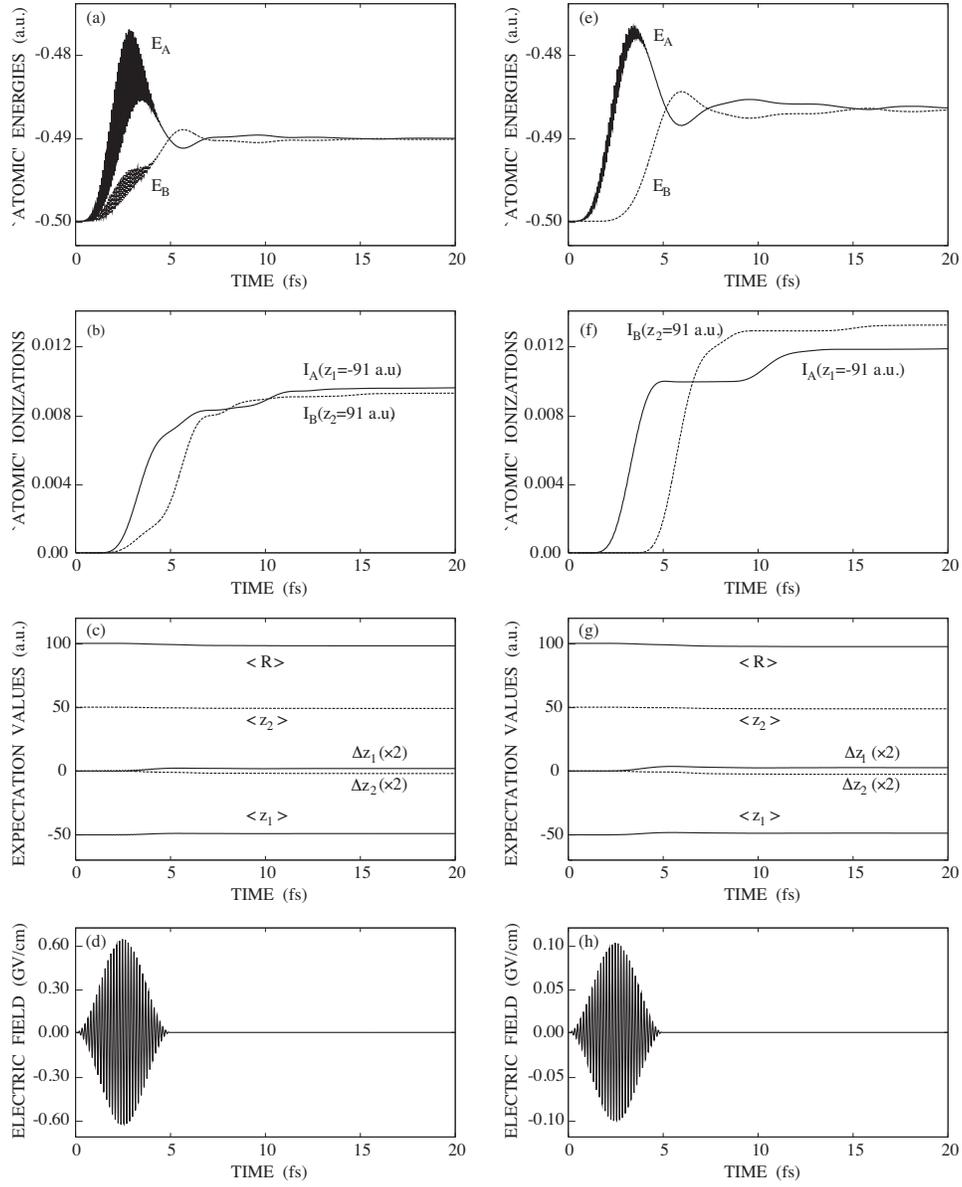}
 \end{center}
 \caption{Quantum dynamics of H-H excited from an unentangled direct-product
 initial state by spatially shaped laser fields: 
 a Gaussian spatial envelope (a-d) and a narrow spatial envelope (e-f).
 (a) and (e) - the energy transfer; (b) and (f) - `atomic' ionizations
 [ionization probabilities ${\rm I}_{\rm A}(z_{1}=91\,{\rm a.u.})$ and 
 ${\rm I}_{\rm B}(z_{2}=-91\,{\rm a.u.})$ are of the order of 10$^{-6}$ and are not shown] ;
 (c) and (g) - expectation values  $\langle R \rangle$, $\langle z_{1} \rangle$ 
 and $\langle z_{2} \rangle$ (deviations of the expectation values, $\Delta z_{1}$ 
 and $\Delta z_{2}$, defined by Eqs.~(\ref{E-Zdeviations}) 
 are scaled up by a factor of two); 
 (d) and (h) - local effective laser fields ${\cal E}^{{\rm A}}(t)$
 [Eqs.~(\ref{E-Eampl}) and (\ref{E-Etime})] acting on atom {\rm A}.}
 \end{figure*}

 In order to study the energy transfer from {\rm A} to {\rm B},
 the respective `atomic' energies, $E_{\rm A}(t)$ and $E_{\rm B}(t)$, 
 have been defined  
 on the basis of Eqs.~(\ref{E-3}) and (\ref{E-4}) 
 as follows: 
 \begin{displaymath}
 E_{\rm A}(t) = \biggl< \Psi (t) \bigg|
 -\frac{1}{2m_{\rm p}}\frac{\partial^{2}}{\partial R^{2}} 
 + \frac{1}{2} V_{\rm p p}(R) 
  + \frac{1}{2} V_{\rm e e}(z_{1},z_{2})
 \end{displaymath}
 \begin{equation}
  -\frac{1}{2 \mu_{\rm e}}\frac{\partial^{2}}{\partial z_{1}^{2}} 
 - \sum _{k=1}^{2} \frac{1}{\sqrt{(z_{k}+R/2)^{2}+\beta}}
  \bigg| \Psi (t) \biggr> 
  \label{E-EnA-DP}
 \end{equation}
 for atom {\rm A} and similarly,
 \begin{displaymath}
 E_{\rm B}(t) = \biggl< \Psi (t) \bigg|
 -\frac{1}{2m_{\rm p}}\frac{\partial^{2}}{\partial R^{2}} 
 + \frac{1}{2} V_{\rm p p}(R) 
  + \frac{1}{2} V_{\rm e e}(z_{1},z_{2})
 \end{displaymath}
 \begin{equation}
  -\frac{1}{2 \mu_{\rm e}}\frac{\partial^{2}}{\partial z_{2}^{2}} 
 - \sum _{k=1}^{2} \frac{1}{\sqrt{(z_{k}-R/2)^{2}+\beta}}
  \bigg| \Psi (t) \biggr>,
  \label{E-EnB-DP}
 \end{equation}
 for atom {\rm B}. 
 From Eqs.~(\ref{E-EnA-DP}) and (\ref{E-EnB-DP}) we see
 that the kinetic energy of nuclear motion and the potential energies 
 of the proton-proton and the electron-electron interaction are assumed
 to be equally shared between atoms~{\rm A} and {\rm B}. 
 The kinetic energy of electron $e_{1}$ 
 and the energy of Coulombic interaction of both electrons, $e_{1}$ and $e_{2}$,
 with proton $p_{\rm A}$
 are entirely assigned to atom {\rm A}.
 Similarly, the kinetic energy of electron $e_{2}$
 and the energy of Coulombic interaction of both electrons
 with proton $p_{\rm B}$
 are entirely assigned to atom {\rm B}.
 The sum of `atomic' energies 
 always gives the correct total energy of the entire H-H system.
 Notice, in particular, that the fact that the kinetic energy of electron
 $e_{1}/e_{2}$ is entirely assigned to atom~{\rm A}/{\rm B} 
 corresponds to the initial electron probabilities $P(z_{1})$/$P(z_{2})$
 in the unentangled atomic state [Fig.~1(a)]. 
 Indeed, electron $e_{1}$ is localized in the vicinity of proton $p_{\rm A}$ of atom~{\rm A}
 and electron $e_{2}$ is localized in the vicinity of proton $p_{\rm B}$ of atom~{\rm B}.
 Therefore, if e.g. the laser pulse with a narrow spatial shape [Fig.~1(b)] is used
 to excite the extended H-H system, only the `atomic' energy $E_{\rm A}(t)$ will increase,
 while $E_{\rm B}(t)$ will not be affected.

 The left panel of Fig.~2 presents the dynamics of the atomic H-H system
 excited by the laser pulse  with the Gaussian spatial envelope 
 \begin{equation}
 F_{\rm G}(z)=\exp\{-[(z-z_{0})/\lambda]^{2}\},
  \label{E-G}
 \end{equation}
 where $\lambda =861$~a.u. (45.56~nm)
 and the laser field is assumed to be focused 
 onto a spot centered at $z_{0}$,
 where $z_{0}=-1.5\lambda$ [$z_{0}=$-1291.5~a.u. (-68.34~nm)]: cf. Fig.~1(b).
 The laser pulse parameters are:
 $\omega =1.0$~a.u., $t_{p}=5$~fs, and ${\cal E}_{0}=1.0$~a.u.
 (${\cal E}_{0}=5.14\times10^{9}$~V/cm, intensity $I_{0}^{\rm A}=3.5\times 10^{16}$~W/cm$^{2}$).
 Accordingly, 
 the spatial envelope function $F(z)$ in the interaction Hamiltonian of Eq.~(\ref{E-6}) 
 is set equal to $F_{\rm G}(z)$ of Eq.~(\ref{E-G}), 
 and the effective-field amplitudes
 for atoms {\rm A} and {\rm B} defined by Eq.~(\ref{E-Eampl}) are:
 ${\cal E}_{0}^{\rm{\rm A}}=0.125$~a.u. 
 (intensity $I_{0}^{\rm A}=5.48\times 10^{14}$~W/cm$^{2}$)
 and
 ${\cal E}_{0}^{\rm{\rm B}}=0.088$~a.u.,
 ($I_{0}^{\rm B}=2.72\times 10^{14}$~W/cm$^{2}$),
 implying a dominant excitation of atom~{\rm A}, which is clearly observed in Fig.~2(a).
 
 From Fig.~2(a) we see that the `atomic' energy $E_{\rm A}(t)$ 
 is controlled by the applied laser pulse:
 it increases in the first half of the pulse and decreases to the end of the pulse.
 In contrast, the `atomic' energy $E_{\rm B}(t)$ does not follow the applied laser pulse:
 it slowly increases up to the end
 of the laser pulse and even exceeds $E_{\rm A}(t)$ shortly after 
 the end of the pulse. 
 A similar behavior is found for the `atomic' ionizations 
 ${\rm I}_{\rm A}$ and ${\rm I}_{\rm B}$ presented in Fig.~2(b):
 while the laser-induced ionization ${\rm I}_{\rm A}$ rises fast in the second half of the pulse,
 the ionization probability ${\rm I}_{\rm B}$ sharply rises after the end of the pulse.
 Such a behavior will be referred to below as a `sequential' ionization,
 implying a delayed interaction due to LIET. 
 The time-dependent expectation values $\langle R(t) \rangle$,
 $\langle z_{1}(t) \rangle$ and $\langle z_{2}(t) \rangle$ 
 are presented  in Fig.~2(c).
 The time-dependent deviations of the expectation values
 $\langle z_{1}(t) \rangle$ and $\langle z_{2}(t) \rangle$,
 \begin{displaymath}
 \Delta z_{1}(t)=\langle z_{1}(t) \rangle-\langle z_{1}(t=0) \rangle,
 \end{displaymath}
 \begin{equation}
 \Delta z_{2}(t)=\langle z_{2}(t) \rangle-\langle z_{2}(t=0) \rangle,
  \label{E-Zdeviations}
 \end{equation}
 plotted in Fig.~2(c) are scaled up by a factor of two.
 We see from Fig.~2(c) that $\Delta z_{1}(t) > 0$,
 implying attraction of electron $e_{1}$ by proton $p_{\rm B}$.
 In contrast, $\Delta z_{2}(t) < 0$,
 implying attraction of electron $e_{2}$ by proton $p_{\rm A}$.
 It is also seen from Figs.~2(a), 2(b) and 2(c) that
 both the energy transfer from {\rm A} to {\rm B} and the `sequential' ionization of {\rm B}
 in the positive direction of the $z_{2}$-axis correlate with a small decrease
 of the spatial separation of electrons,
 $|\langle z_{2} \rangle - \langle z_{1} \rangle |$,  
 such that the electron-electron repulsion (EER) becomes effective.
 
 In order to clarify these findings further, we have performed a similar numerical simulation
 with a narrow spatial envelope of the laser pulse acting practically only on atom~{\rm A} 
 of the entire H-H system. 
 Specifically, the following sin$^{2}$-type spatial envelope of the laser pulse was used for 
 $z=z_{1}$ and $z=z_{2}$:
 \begin{equation}
  F_{\rm N}(z)=\sin^{2}\biggl[\frac{\pi(z-z_{a})}{z_{b}-z_{a}}\biggr], \; \; \; z_{a} \leq z \leq z_{b}, 
  \label{E-N}
 \end{equation}
 where $z_{a}=-60$~a.u., $z_{b}=-40$~a.u., and $F_{\rm N}(z)=0$ otherwise,
 as illustrated in Fig.~1(b).
 Accordingly, 
 the spatial envelope function $F(z)$ in Eq.~(\ref{E-6}) 
 is set equal to $F_{\rm N}(z)$ of Eq.~(\ref{E-N}). 
 The amplitude of the pulse, ${\cal E}_{0}=0.02$~a.u.
 ($I_{0}=1.4\times 10^{13}$~W/cm$^{2}$), 
 is chosen such that
 the `atomic' energy $E_{\rm A}(t)$ gained in the field 
 with the narrow spatial envelope [Fig.~2(e)] 
 is close to that gained with the Gaussian spatial envelope [Fig.~2(a)].
 Note that, defined by Eq.~(\ref{E-Etime}), the effective-field strength ${\cal E}^{\rm A}(t)$ 
 of the laser pulse with the narrow spatial envelope
 [Figs.~1(b) and 2(h)]  
 is much smaller than
 that of the laser pulse with the Gaussian spatial envelope [Figs.~1(b) and 2(d)], 
 but the energy transfer from {\rm A} to {\rm B} is by about 30\% more efficient [Fig.~2(e)].
 Also note that, due to the effective-field amplitude acting on atom~{\rm B}
 in the current case of the narrow spatial envelope of the pulse
 is ${\cal E}_{0}^{\rm B}=0$,
 the efficient energy transfer to atom~{\rm B}
 is attributed entirely
 to the electron-electron repulsion $V_{\rm e e}(z_{1},z_{2})$
 and electron-proton attraction $V_{\rm e p}(z_{1(2)},R)$
 in the vicinity of atom~{\rm B} ($z_{1(2)} = z_{\rm B}$).
 This also applies to the ionization of atom~{\rm B} 
 in the positive direction of the $z_{2}$-axis [Fig.~2(f)], 
 which is not only by about 30\% more efficient than that induced
 by the laser pulse with the broad Gaussian spatial envelope [Fig.~2(b)],
 but ${\rm I}_{\rm B}$ is even larger than ${\rm I}_{\rm A}$ at $t > 7$~fs [Fig.~2(f)], 
 implying that the `sequential' ionization ${\rm I}_{\rm B}$, 
 induced by EER,
 is more efficient than the laser-induced ionization ${\rm I}_{\rm A}$.

 Taking into account that in the unentangled direct-product state (\ref{ini-DP}), 
 electrons $e_{1}$ and $e_{2}$ are well localized on the $z$-axis [Fig.~1(a)],
 one can assume that the time-dependent expectation values 
 $\langle z_{1}(t) \rangle$ and $\langle z_{2}(t) \rangle$
 represent the positions of electrons.
 A closer look at the data presented in Figs.~2(c) and 2(g)  
 for $\langle z_{1}(t) \rangle$ and $\langle z_{2}(t) \rangle$ shows
 that the two electrons of H-H do not approach each other closer than 98~a.u. (5.18~nm).
 Therefore, at a first glance, the energy transfer from {\rm A} to {\rm B} 
 and the `sequential' ionization of {\rm B}
 take place,
 similarly to ICD
 \cite{Cederbaum:97prl.ICD,%
 Marburger:03prl.ICD,%
 Jahnke:04prl.ICD,%
 Morishita:06prl.ICD,%
 Lablanquie:07jcp.ICD,%
 Jahnke:10NaturePhys.ICD,%
 Havermeier:10prl.ICD,%
 Sisourat:10NatPhys.ICD},
 without any noticeable overlap of the respective electronic wave functions.
 Nevertheless, a closer look at spatial distributions of the electronic
 wave functions illustrated in Fig.~3 
 reveals that EER is in fact
 a short-range process which takes place in the vicinity of atom~{\rm B}
 at $z_{\rm B}=50$~a.u..
 Here, the Coulombic attraction of electron $e_{1}$ by proton $p_{\rm B}$
 effectively localizes electron $e_1$.
 Accordingly, the overlap of the electronic wave functions
 has a sharp narrow maximum at $z_{\rm B}=50$~a.u. (Fig.~3),
 which results in a strong EER. 

 Electron probabilities $P(z_{1})$ and $P(z_{2})$ presented in Fig.~3
 correspond to the end of the 5~fs laser pulse with the narrow spatial envelope
 defined by Eq.~(\ref{E-N}). This pulse 
 excites practically only electron $e_{1}$ in the vicinity of $z_{\rm A}=-50$~a.u., 
 therefore the probability distribution $P(z_{1})$ at $t=5$~fs
 is broadened in comparison to the initial one [Fig.~1(a)]
 in both $z_{1}<z_{\rm A}$ and $z_{1}>z_{\rm A}$ directions. 
 The laser-induced extension of $P(z_{1})$ into the domain of  $z_{1}<z_{\rm A}$ 
 gives rise to the ionization ${\rm I}_{\rm A}$ on the negative direction of the $z_{1}$-axis.
 At $z_{1}\leq -91$~a.u., 
 the wave function is absorbed by the imaginary optical potential.
 In contrast, at $z_{1}>z_{\rm A}$, the laser-driven electron $e_{1}$ reaches
 the domain $z_{1}>0$ where it is attracted and accelerated by proton $p_{\rm B}$,
 localized at $z_{\rm B}=50$~a.u..
 Due to the electron-proton Coulombic attraction, the probability $P(z_{1})$ has its local
 maximum at $z_{1}=z_{\rm B}$, corresponding to LIET of Ref. 
 \cite{Bandr:07prl.HpH2pBO}.
 Simultaneously, when electron $e_{1}$ approaches the domain of $z_{\rm B}=50$~a.u., 
 where the initial probability $P(z_{2})$ of electron $e_{2}$ has the global maximum [Fig.~1(a)], 
 the electron-electron Coulombic repulsion becomes very strong. 
 Therefore, the probability $P(z_{2})$ is extended into the domain of $z_{2}>z_{\rm B}$ (Fig.~3),
 giving rise to the ionization ${\rm I}_{\rm B}$ in the positive direction of the $z_{2}$-axis.
 Finally, the wave function is absorbed by the imaginary optical potential
 at $z_{2}\geq 91$~a.u..
 Note that in the current case of a narrow spatial envelope of the laser field [Fig.~1(b)]
 one can clearly distinguish between the laser-induced ionization 
 on the negative direction of the $z_{1}$-axis,
 represented in Fig.~2(f) 
 by ${\rm I}_{\rm A}(z_{1}=-91\,{\rm a.u.})$,
 and the `sequential' EER-induced ionization on the positive direction of the $z_{2}$-axis,
 represented in Fig.~2(f) by ${\rm I}_{\rm B}(z_{2}=91\,{\rm a.u.})$.
 Also note that the ionization probabilities 
 ${\rm I}_{\rm A}(z_{1}=91\,{\rm a.u.})$
 and
 ${\rm I}_{\rm B}(z_{2}=-91\,{\rm a.u.})$
 are less than 10$^{-6}$ at $t=5$~fs.

 \begin{figure}[t]
 \includegraphics*[width=16pc]{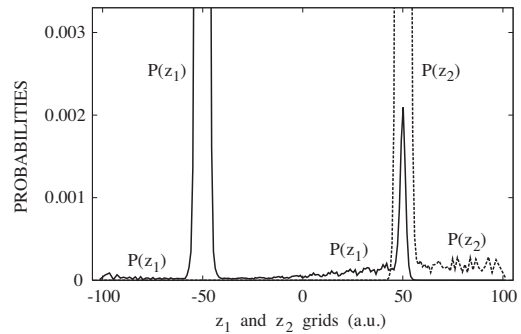}
 \caption{Electron probabilities $P(z_{1})$ (solid line)
 and $P(z_{2})$ (dashed line) 
 at the end of the 5~fs laser pulse with a narrow spatial envelope
 [Eq.~(\ref{E-N}) and Fig.~1(b)] which excites only electron $e_{1}$ 
 (coordinate $z_{1}$) in the domain of atom~{\rm A}.}
 \end{figure}

 \begin{figure}[b]
 \includegraphics*[width=16pc]{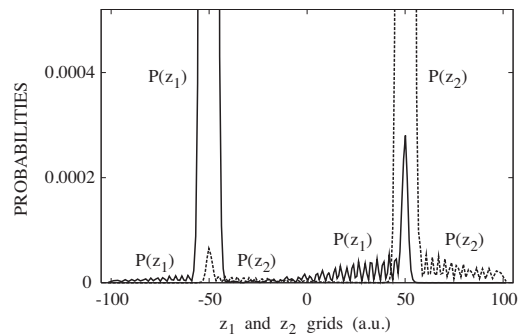}
 \caption{Electron probabilities $P(z_{1})$ (solid line)
 and $P(z_{2})$ (dashed line) 
 at 2.5~fs after the end of the 5~fs laser pulse with a Gaussian spatial envelope
 which excites both electrons, with the excitation of electron $e_{1}$ 
 (coordinate $z_{1}$) being dominant in comparison to that of electron $e_{2}$
 (coordinate $z_{2}$).}
 \end{figure}
 
 In the case of the broad Gaussian spatial envelope of the laser pulse
 centered at $z_{0}=-1291.5$~a.u. (-68.34~nm), 
 both electrons, $e_{1}$ and $e_{2}$, are excited by the laser field
 and therefore ionization in both positive and negative directions
 of the $z$-axis is induced by both the laser field 
 and subsequently by EER.
 The electron probabilities $P(z_{1})$ (solid line) and $P(z_{2})$ (dashed line)
 for the Gaussian spatial envelope of the 5~fs laser pulse are plotted in Fig.~4 at $t=7.5$~fs,
 when both ionization probabilities plotted in Fig.~2(b) approach their maximum values.
 Although both electrons are excited by the laser field, the excitation of
 electron $e_{1}$ in the domain of atom~{\rm A} is about 1.4 times stronger than the excitation of 
 electron $e_{2}$ in the domain of atom~{\rm B}. 
 Accordingly, the local maximum of $P(z_{1})$ 
 at $z_{\rm B}=50$~a.u. (Fig.~4) is almost 4 times higher than the local
 maximum of $P(z_{2})$ at $z_{\rm A}=-50$~a.u. 
 and therefore the EER-induced ionization 
 in the positive direction of the $z_{2}$-axis is stronger than 
 the EER-induced ionization in the negative direction of the $z_{1}$-axis.
 In contrast, the laser-induced ionization in the negative direction of the $z_{1}$-axis
 is stronger than that in the positive direction of the $z_{2}$-axis. 
 Indeed, it is seen from Fig.~2(b) that at the end of the 5~fs laser pulse,
 the ionization probability ${\rm I}_{\rm B}$ is more than twice smaller than 
 ${\rm I}_{\rm A}$, while ${\rm I}_{\rm B} \approx {\rm I}_{\rm A}$ at $t>$10~fs.
 
 It is seen from Fig.~4 that both local maxima, $P(z_{1})$ at $z_{1}=-50$~a.u.
 and $P(z_{2})$ at $z_{2}=50$~a.u., are significantly smaller than the local maximum of $P(z_{1})$ 
 at $z_{1}=-50$~a.u. produced by the laser pulse with a narrow spatial envelope (see Fig.~3).
 Therefore the EER-induced ionization in the case of the broad Gaussian spatial envelope 
 of the laser pulse [Fig.~2(b)] is smaller than 
 that in the case of the narrow spatial envelope of the pulse [Fig.~2(f)].

 It can be concluded from Fig.~4 that
 in the general case, when both distant atoms are excited by the laser field,
 energy is transferred from {\rm A} to {\rm B}
 and from {\rm B} to {\rm A}, and EER-induced ionization occurs 
 in both {\rm A} and {\rm B} parts of H-H along with the laser-induced ionization. 
 If, for example, the laser pulse has a wide Gaussian spatial envelope and/or is centered at $z=0$,
 the mutual energy transfers and the EER-induced ionization probabilities are substantial
 for both distant atoms {\rm A} and {\rm B} even at a large internuclear separation.

 Finally we note that the appearance of sharp local maxima of $P(z_{1})$ at $z_{1}=50$~a.u.
 and of $P(z_{2})$ at $z_{2}=-50$~a.u. (see Figs.~3 and 4) 
 confirm spreading, delocalization, and non-factorization of the initially factorized 
 and well localized on the $z$-axis wave function of H-H [Fig.~1(a)]
 and can therefore 
 \cite{Eberly:04pra.Entang} 
 be treated as the emergence of long-range entanglement,
 or quantum non-local connection, at $R=100$~a.u..
 Recently this long-range entanglement attracted considerable interest both in theory and experiment
 \cite{BandrShon:pra.VPA.Entang,%
 ShapiroM:03pra.EntangTransfer,%
 BurkardDiVi:99prb.SpinEntang,%
 Wilk:2010prl.EntangExp,%
 IshizakiFleming:2010NewJPhys.BiomolEntangTheor}.

 \section{Excitation of H-H from an entangled initial state} 

 The H-H molecular system represents an elongated
 configuration of the H$_{2}$ molecule. 
 Therefore, its initial electronic ground state is entangled via exchange 
 \cite{Havermeier:10prl.ICD,%
 Bandr:08prl.H2d3,%
 Wilk:2010prl.EntangExp}.
 The spatial part of the initial entangled ground-state wave function 
 of H-H for a singlet electronic state is given by
 \begin{displaymath}
 \Psi(R,z_{1},z_{2},t=0)=[\Psi_{1S_{\rm A}}(z_{1}) \Psi_{1S_{\rm B}}(z_{2}) 
 \end{displaymath}
 \begin{equation}
 + \Psi_{1S_{\rm B}}(z_{1}) \Psi_{1S_{\rm A}}(z_{2})] \Psi_{\rm G}(R),
  \label{ini-ENp}
 \end{equation}
 where $\Psi_{1S_{\rm A,B}}(z_{1,2})$ are defined by Eq.~(\ref{ini-DP})   
 and $\Psi_{\rm G}(R)$ is a proton Gaussian function centered at $R=100$~a.u..
 Imaginary time propagations have been performed 
 with the complete system Hamiltonian (\ref{E-3}).
 In the entangled initial state of the H-H system,
 electron probabilities $P(z_{1})$ and $P(z_{2})$ are identical
 to each other.

 For the simulations with this entangled initial state (\ref{ini-ENp}), 
 the `atomic' energies
 $E_{\rm A}(t)$ and $E_{\rm B}(t)$ are defined on the basis 
 of Eqs.~(\ref{E-3}) and (\ref{E-4}) as follows:
 \begin{displaymath}
 E_{\rm A}(t) = \biggl< \Psi (t) \bigg|
 -\frac{1}{2m_{\rm p}}\frac{\partial^{2}}{\partial R^{2}} 
  - \frac{1}{2}\sum _{k=1}^{2} \frac{1}{2 \mu_{\rm e}}\frac{\partial^{2}}{\partial z_{k}^{2}} 
 + \frac{1}{2} V_{\rm p p}(R) 
 \end{displaymath}
 \begin{equation}
  + \frac{1}{2} V_{\rm e e}(z_{1},z_{2}) 
 - \sum _{k=1}^{2} \frac{1}{\sqrt{(z_{k}+R/2)^{2}+\beta}}
  \bigg| \Psi (t) \biggr> 
  \label{E-EnA-ENp}
 \end{equation}
 for atom {\rm A}, and
 \begin{displaymath}
 E_{\rm B}(t) = \biggl< \Psi (t) \bigg|
 - \frac{1}{2m_{\rm p}} \frac{\partial^{2}}{\partial R^{2}} 
  - \frac{1}{2}\sum _{k=1}^{2} \frac{1}{2 \mu_{\rm e}} \frac{\partial^{2}}{\partial z_{k}^{2}} 
 + \frac{1}{2} V_{\rm p p}(R) 
 \end{displaymath}
 \begin{equation}
  + \frac{1}{2} V_{\rm e e}(z_{1},z_{2}) 
 - \sum _{k=1}^{2} \frac{1}{\sqrt{(z_{k}-R/2)^{2}+\beta}}
  \bigg| \Psi (t) \biggr> 
  \label{E-EnB-ENp}
 \end{equation}
 for atom {\rm B}. 
 From Eqs.~(\ref{E-EnA-ENp}) and (\ref{E-EnB-ENp}) we see
 that the kinetic energies of nuclear and electronic motion
 as well as the potential energies of
 the proton-proton and the electron-electron Coulombic interaction are assumed
 to be equally shared between atoms {\rm A} and {\rm B}. 
 The energy of Coulombic interaction of both electrons with
 proton $p_{\rm A}$ is assigned to atom {\rm A}, while
 the energy of Coulombic interaction of both electrons with
 proton $p_{\rm B}$ is assigned to atom {\rm B}.
 The sum of these `atomic' energies 
 always gives the correct total energy of the entire H-H system.
 The choice of the electron kinetic energies in Eqs.~(\ref{E-EnA-ENp}) and (\ref{E-EnB-ENp})
 corresponds to the initial electron probabilities $P(z_{1})$ and $P(z_{2})$ in the
 entangled molecular state. Both electrons $e_{1}$ and $e_{2}$ are localized with the 50\%
 probability in the vicinity of proton $p_{\rm A}$ of atom~{\rm A} and if the extended 
 H-H system is excited e.g. by the narrowly shaped laser pulse of Fig.~1(b), 
 both electrons give rise to the `atomic' energy $E_{\rm A}(t)$, each with 50\% probability.

 \begin{figure*}[t]
 \begin{center}
 \includegraphics*[width=30pc]{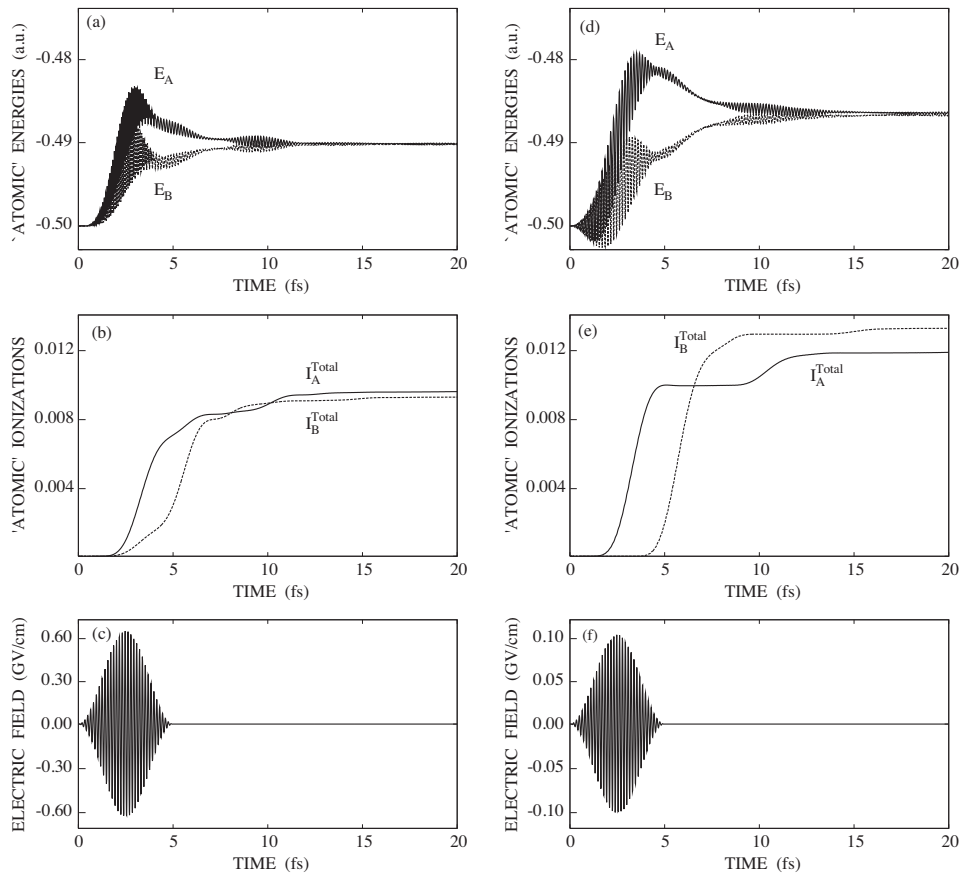}
 \end{center}
 \caption{Quantum dynamics of H-H excited from the entangled
 initial state (\ref{ini-ENp}) by spatially shaped laser fields [Fig.~1(b)]: 
 a broad Gaussian spatial envelope (left panel) and a narrow spatial envelope (right panel).
 (a) and (d) - the energy transfer; (b) and (e) - `atomic' ionizations;
 [the total ionization probabilities ${\rm I}_{\rm A}^{\rm Total}$
 and ${\rm I}_{\rm B}^{\rm Total}$ are defined by Eq.~(\ref{E-Flexes})];
 (c) and (f) - local effective laser fields ${\cal E}^{{\rm A}}(t)$
 [Eqs.~(\ref{E-Eampl}) and (\ref{E-Etime})] acting on atom {\rm A}.}
 \end{figure*}

 The quantum dynamics of H-H excited from the entangled initial state (\ref{ini-ENp})
 by spatially shaped laser pulses is presented in Fig.~5.
 The left panel corresponds to the broad Gaussian spatial envelope 
 centered at $z_{0}=-1291.5$~a.u. (-68.34~nm) and the right panel
 corresponds to the narrow spatial envelope centered 
 at $z_{\rm A}=-50$~a.u. (-2.65~nm), see Fig.~1(b). 
 For the sake of comparison, the laser fields used are the same as in the previous
 case of the unentangled direct-product initial state [see Figs.~1(b) and 2]. 
 Specifically, the Gaussian spatial envelope is defined by Eq.~(\ref{E-G}),
 the narrow spatial envelope is defined by Eq.~(\ref{E-N}),
 the carrier frequency of the laser pulse with the sin$^{2}$-type temporal envelope of Eq.~(\ref{E-Etime})
 is $\omega =1.0$~a.u., and the pulse duration at the base is $t_{p}=5$~fs.

 From the comparison of the quantum dynamics of the unentangled state [Figs.~2(a) and 2(e)]
 to that of the entangled state [Figs.~5(a) and 5(d)]
 the following observations are made.

 (i) The overall energy $\Delta E_{\rm B}$ transferred on a long timescale of $t=20$~fs
 from atom~{\rm A} to atom~{\rm B} 
 in the entangled state is very similar 
 to that transferred from {\rm A} to {\rm B} in the unentangled direct-product state: 
 $\Delta E_{\rm B}\approx 0.01$~a.u. for the broad Gaussian spatial envelope, and
 $\Delta E_{\rm B}\approx 0.013$~a.u. for the narrow spatial envelope.

 (ii) The maximum `atomic' energy $E_{\rm A}(t)$ gained during the laser pulse by atom~{\rm A} 
 in the entangled state is smaller than that gained 
 in the unentangled direct-product state.
 In contrast, the maximum `atomic' energy $E_{\rm B}(t)$ gained during the pulse by atom~{\rm B}
 in the entangled state is substantially larger than that gained 
 in the unentangled direct-product state.

 (iii) Moreover, in the case of the entangled initial state, 
 the `atomic' energy $E_{\rm B}(t)$ 
 is controlled
 by the laser pulse similarly to $E_{\rm A}(t)$, even when only atom~{\rm A} is excited 
 by the laser pulse with a narrow spatial envelope: 
 the energy $E_{\rm B}(t)$ increases in the first half of the laser pulse
 and decreases at the end of the pulse, in contrast to the case of the unentangled direct-product initial state.
 This observation implies that the changes made by the applied laser field 
 to the entangled wave function in the domain of atom~{\rm A}
 at $z_{1,2}\approx -50$~a.u. result in simultaneous changes in atom~{\rm B} at $z_{1,2}\approx 50$~a.u.
 due to the symmetry of the wave function by exchange.
 Such an entangled behaviour is very important 
 for a long-range quantum communication among distant quantum systems
 \cite{BurkardDiVi:99prb.SpinEntang}.

 (iv) After the end of the laser pulse, the `atomic' energies $E_{\rm A}(t)$ and $E_{\rm B}(t)$
 demonstrate out-of-phase oscillations: slow oscillations in the case 
 of the unentangled direct-product initial state [Figs.~2(a) and 2(e)]
 and slow oscillations modulated with very fast ones in the case of the entangled initial state [Figs.~5(a) and 5(d)].
 The amplitudes of both slow and fast oscillations of $E_{\rm A}(t)$ and $E_{\rm B}(t)$ decrease
 with time, which may indicate the formation of a quasi-stable configuration of the excited H-H system.

 \begin{figure}[t]
 \begin{center}
 \includegraphics*[width=16pc]{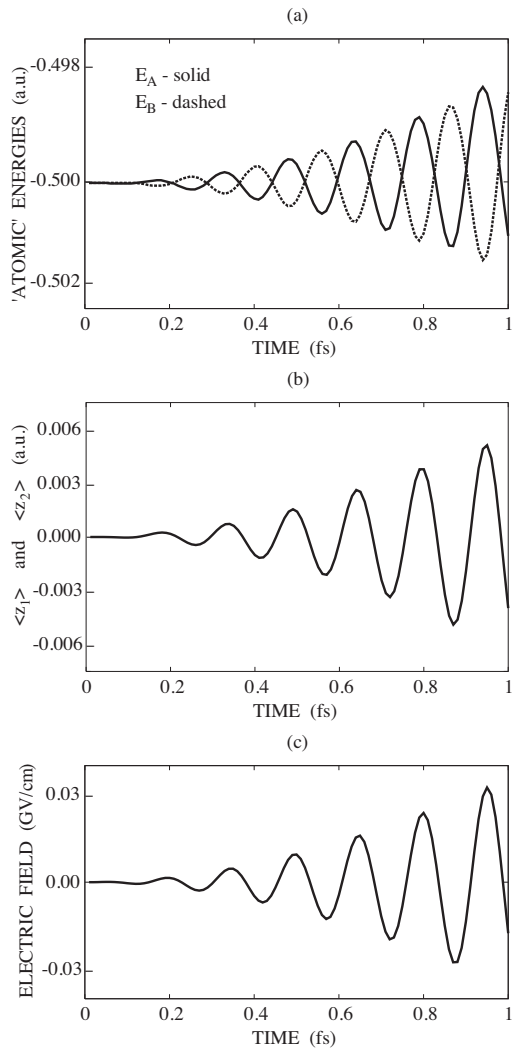}
 \end{center}
 \caption{The initial (1~fs) stage of excitation of H-H from the entangled initial state [Eq.~(\ref{ini-ENp})]
 by the laser pulse with a narrow spatial envelope [Eq.~(\ref{E-N})] acting on atom~{\rm A}.
 (a) - time-dependent `atomic' energies $E_{\rm A}(t)$ (solid line)
 and $E_{\rm B}(t)$ (dashed line);
 (b) - electron-field in-phase following;
 (c) - the laser field acting on atom~{\rm A}.} 
 \end{figure}

 It is instructive, before proceeding, to consider the laser-driven dynamics of the entangled H-H system
 in the initial stage of its excitation by the laser pulse with a narrow spatial envelope (\ref{E-N})
 which excites electrons $e_{1}$ and $e_{2}$ only in the domain of atom~{\rm A}.
 In Fig.~6, the laser-driven dynamics of entangled H-H is presented
 on the timescale of 1~fs.
 Time-dependent `atomic' energies, $E_{\rm A}(t)$ and $E_{\rm B}(t)$, 
 and expectation values of electronic coordinates, 
 $\langle z_{1}(t) \rangle$ and $\langle z_{2}(t) \rangle$,
 are shown in Figs.~6(a) and 6(b) respectively.
 The time-dependent laser field is shown in Fig.~6(c). 

 It is seen from Fig.~6(a) that the `atomic' energies $E_{\rm A}(t)$ and $E_{\rm B}(t)$ 
 oscillate out-of-phase with respect to each other,
 with $E_{\rm A}(t)$ being in-phase and $E_{\rm B}(t)$ being out-of phase 
 with the applied laser field [Fig.~6(c)]. 

 As seen from Figs.~6(b) and 6(c) the electrons also
 follow the applied laser field. 
 Note that expectation values $\langle z_{1}(t) \rangle$ and $\langle z_{2}(t) \rangle$
 in the entangled H-H system are identical and not distinguishable in Fig.~6(b).
 A perfect electron-field following at the laser carrier frequency 
 being as high as 1~a.u. is very interesting.
 Previously, the electron-field following on the level of expectation values
 of electronic coordinates have been explored only in the infrared 
 \cite{Paramon:05cpl.HH-HD.07cp.HH-HD-Muon}
 and near-infrared
 \cite{Kono:04cp.H2p.nearIR}
 domains of the laser carrier frequency.
 The electron-field following at high laser frequencies is reminiscent of the well-known
 recollision model of Corkum
 \cite{Corkum:93prl.Recollision}.
 However, an important feature of Fig.~6(b) is that electrons follow the applied laser field
 in-phase, while according to the theoretical model used in
 \cite{Corkum:93prl.Recollision}
 and numerical results of
 \cite{Paramon:05cpl.HH-HD.07cp.HH-HD-Muon}
 electrons follow the field out-of-phase: 
 $\langle z(t) \rangle$ decreases when electric-field strength ${\cal E}(t)$ increases.
 A detailed study of the electron dynamics showed that,
 probably due to their finite albeit very small mass,
 the electron do not react to the first half-cycle of the applied field at $\omega =1$~a.u.
 and follows the field in-phase at $t>0.15$~fs.
 On the other hand, the out-of-phase electron-field following take place at $\omega < 0.1$~a.u..
 Similar results have been obtained for the unentangled H-H system excited by the laser pulse
 with a narrow spatial envelope: 
 the electron $e_{1}$ of atom~{\rm A} follows the applied field in-phase at $\omega =1$~a.u.
 and out-of-phase at $\omega < 0.1$~a.u..
 These results can be explained as follows. At $\omega < 0.1$~a.u., the electron is in the 
 ground state, the polarizability is negative and, therefore, 
 the electron follows the applied laser field out-of-phase.
 In contrast, at $\omega =1$~a.u., the electron is well above the 
 excited state, the polarizability changes the sign, and electron follows the field in-phase.
 Similar behaviour has been observed in a recent work
 \cite{Bandrauk:09pra.Recollision}
 for molecular ion H$_{2}^{+}$ excited at wavelength $\lambda =$800~nm ($\omega=0.057$~a.u.).
 At fixed $R=2$~a.u., the electron is in the ground state and follows the field out-of-phase,
 whereas at $R=7$~a.u., the laser carrier frequency $\omega=0.057$~a.u. is larger than
 the energy difference $E_{1s\sigma_{u}}-E_{1s\sigma_{g}}$, and the electron follows the field
 in-phase.

 Coming back to the laser-driven dynamics of the entangled H-H system 
 on the long timescale of 20~fs (Fig.~5),
 we note that ionization of atom~{\rm B} starts only in the second half of the
 laser pulse [see Figs.~5(b) and 5(e)].
 Due to the symmetry of the entangled wave function,
 the ionization probabilities of electrons $e_{1}$ and $e_{1}$
 are identical to each other both for the positive and the negative directions
 of the $z$-axes:
 ${\rm I}_{\rm A}(z_{1}=-91\,{\rm a.u.})={\rm I}_{\rm A}(z_{2}=-91\,{\rm a.u.})$
 and
 ${\rm I}_{\rm B}(z_{1}=91\,{\rm a.u.})={\rm I}_{\rm B}(z_{2}=91\,{\rm a.u.})$.
 Therefore, in Figs.~5(b) and 5(e) the total ionization probabilities,
 \begin{displaymath}
 {\rm I}_{\rm A}^{\rm Total}={\rm I}_{\rm A}(z_{1}=-91\,{\rm a.u.})
 +{\rm I}_{\rm A}(z_{2}=-91\,{\rm a.u.}),
 \end{displaymath}
 \begin{equation}
 {\rm I}_{\rm B}^{\rm Total}={\rm I}_{\rm B}(z_{1}=91\,{\rm a.u.})
 +{\rm I}_{\rm B}(z_{2}=91\,{\rm a.u.}),
  \label{E-Flexes}
 \end{equation}
 are plotted.
 From the comparison of Figs.~2(b) and 2(f) to Figs.~5(b) and 5(e)
 one can easily see that the total time-dependent ionization probabilities 
 ${\rm I}_{\rm A}^{\rm Total}$ and ${\rm I}_{\rm B}^{\rm Total}$
 used in the case of the entangled initial state are almost equal (but not identical) 
 to the time-dependent ionization probabilities
 ${\rm I}_{\rm A}(z_{1}=-91\,{\rm a.u.})$ and ${\rm I}_{\rm B}(z_{2}=91\,{\rm a.u.})$
 used in the case of the unentangled direct-product initial state.
 One can conclude therefore that the entanglement of the initial state of H-H,
 including its exchange symmetry, does not change its ionization probability
 in comparison to the non-symmetric unentangled direct-product initial state.

 \begin{figure*}[t]
 \begin{center}
 \includegraphics*[width=31pc]{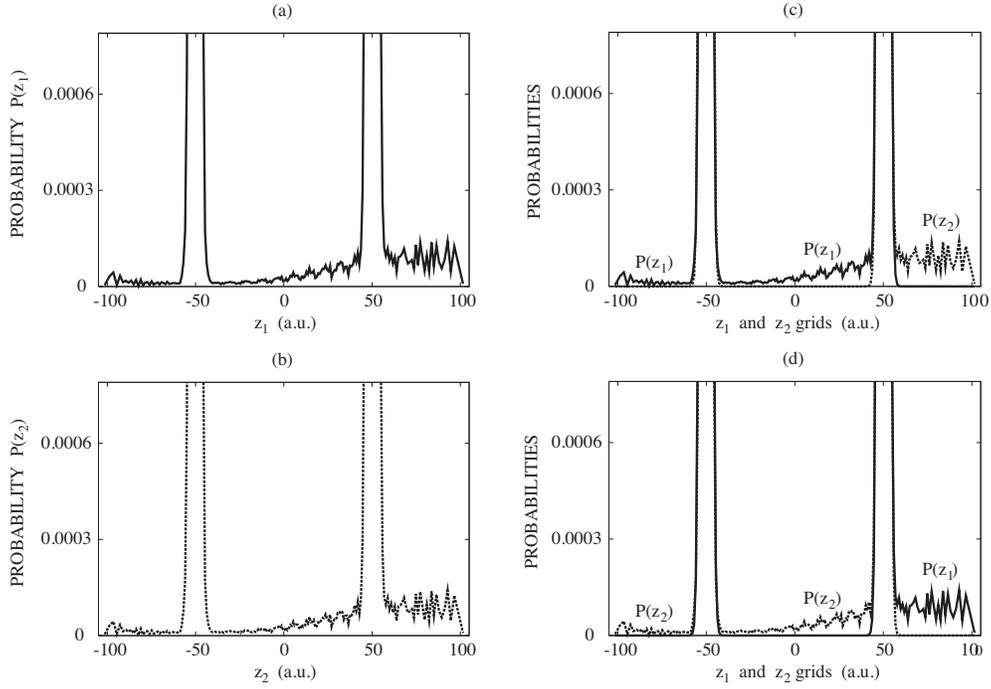}
 \end{center}
 \caption{Excitation of H-H by the laser pulses with the narrow spatial envelopes
 from the entangled initial state:
 electron probabilities $P(z_{1})$ (solid lines) and $P(z_{2})$ (dashed lines)
 at the end of the 5~fs laser pulses.
 (a) and (b) - both electrons $e_{1}$ and $e_{2}$ are excited by the laser pulse 
 in the domain of atom~{\rm A};
 (c) - only electron $e_{1}$ is excited by the laser pulse in the domain of atom~{\rm A};
 (d) - only electron $e_{2}$ is excited by the laser pulse in the domain of atom~{\rm A}.}
 \end{figure*}

 The other consequence of the exchange symmetry of the entangled wave function
 is a very small spatial separation of electrons all over the $z$-grid. 
 The time-dependent expectation values
 $\langle z_{1}(t) \rangle$ and $\langle z_{2}(t) \rangle$ are almost identical,
 both being close to 0 on the timescale of 20~fs. 
 If the perfect symmetry of the wave functions is even slightly changed
 due to the excitation of H-H by the laser field,
 the very small spatial separation of the electrons leads to their strong Coulombic repulsion
 all over the $z$-grid.
 This suggests the reason for the fast out-of-phase oscillations
 of `atomic' energies $E_{\rm A}(t)$ and $E_{\rm B}(t)$ after the end of the laser pulse
 [see observation (iv) above and Figs.~5(a) and 5(d)].
 The fast out-of-phase oscillations of the `atomic' energies
 occur in the case of the entangled initial state and do not occur 
 in the case of the unentangled direct-product initial state [Figs.~2(a) and 2(e)].
 In the case of the direct-product initial state, 
 the minimum difference between $\langle z_{1}(t) \rangle$ and $\langle z_{2}(t) \rangle$
 is about 98~a.u. (5.18~nm) [see Figs.~2(c) and 2(g)], therefore EER is very weak
 all over the $z$-grid, 
 except for the special cases of very sharp overlaps of the electronic wave functions
 in the vicinity of protons $p_{\rm A}$ and $p_{\rm B}$  (see Figs.~3 and 4).

 A very strong EER taking place in the case of the entangled initial state
 should also result in a very efficient EER-induced `sequential' ionization,
 similar to that described in Sec.~III for the unentangled direct-product initial state of H-H. 
 On the other hand, it was shown above that the entanglement of the initial state of H-H
 does not change its ionization probability in comparison to the direct-product initial state.
 Therefore, a closer look at the process of EER-induced ionization in the case of the entangled
 initial state is required. To this end, we performed several model simulations with 
 the entangled initial state excited in the domain of atom~{\rm A} 
 by the laser pulses with narrow spatial envelopes defined by Eq.~(\ref{E-N}) 
 for various model assumptions of the system-field interaction 
 to be specified below.
 The results obtained are presented in Figs.~7 and 8.

 Electron probabilities $P(z_{1})$ and $P(z_{2})$ presented in Figs.~7(a) and 7(b)
 correspond to the end of the 5~fs laser pulse with a narrow spatial envelope [Fig.~1(b)] which
 excites both electrons $e_{1}$ and $e_{2}$ in the vicinity of $z_{\rm A}=-50$~a.u. (atom~{\rm A})
 and does not affect electrons $e_{1}$ and $e_{2}$ 
 in the vicinity of $z_{\rm B}=50$~a.u. (atom~{\rm B}).
 Apparently, electron probabilities $P(z_{1})$ and $P(z_{2})$, 
 presented in Figs.~7(a) and 7(b) at $t=5$~fs,
 are extended in comparison to their initial ones 
 in both $z_{1,2}<z_{\rm A}$ and $z_{1,2}>z_{\rm A}$ directions. 
 The laser-induced extension of $P(z_{1})$ and $P(z_{2})$ into domains $z_{1,2}<z_{\rm A}$ 
 gives rise to the ionization ${\rm I}_{\rm A}$ in the negative directions of the $z_{1,2}$-axes.
 At $z_{1,2}\leq -91$~a.u.,
 the wave function is absorbed by the imaginary optical potentials.
 In contrast, at $z_{1,2}>z_{\rm A}$, the laser-driven electrons $e_{1}$ and $e_{2}$ reach
 the domains $z_{1,2}>0$ where they are attracted and accelerated by proton $p_{\rm B}$,
 localized at $z_{\rm B}=50$~a.u.,
 more and more efficiently.
 Due to the electron-proton Coulomb attraction, electron probabilities 
 $P(z_{1})$ and $P(z_{2})$ increase
 in the vicinity of $z_{1,2}=z_{\rm B}$ and thus destroy the 
 initially perfect symmetry of the entangled wave function, 
 as clearly seen from Figs.~7(a) and 7(b).
 The extension of electron probabilities $P(z_{1})$ and $P(z_{2})$ 
 into the domains of $z_{1,2}>z_{\rm B}$ gives rise to the EER-induced `sequential'
 ionization ${\rm I}_{\rm B}$ in the positive directions of the $z_{1,2}$-axes.
 The only problem to be clarified now is EER in 
 the domain of atom~{\rm B}, because at a first glance one could conclude
 from Figs.~7(a) and 7(b) 
 that the electrons coming from atom~{\rm A} to atom~{\rm B}
 occupy the domains of $z_{1,2}>z_{\rm B}$.
 In order to clarify this issue, we performed two model simulations with 
 a 5~fs laser pulse having narrow spatial envelope, 
 assuming that only one of the two electrons
 is excited by the laser field in the domain of atom~{\rm A}. 
 The results obtained are presented in Figs.~7(c) and 7(d).

 First we assume that only electron $e_{1}$ is excited by the laser field
 in the domain of atom~{\rm A}.
 To this end, we chose the spatial envelope function $F(z)$ 
 in the interaction Hamiltonian of Eq.~(\ref{E-6}) as follows:
 $F(z_{1})=F_{\rm N}(z_{1})$ and $F(z_{2})=0$, 
 where the narrow spatial envelope $F_{\rm N}(z)$ is given by Eq.~(\ref{E-N})
 and illustrated in Fig.~1(b).
 The respective electron probabilities $P(z_{1})$ and $P(z_{2})$
 at the end of the pulse ($t=5$~fs) are shown in Fig.~7(c).
 Secondly, we assume that only electron $e_{2}$ is excited by the laser field
 in the domain of atom~{\rm A}.
 Accordingly, the spatial envelope function $F(z)$ 
 in the interaction Hamiltonian of Eq.~(\ref{E-6}) is chosen as follows:
 $F(z_{1})=0$, and $F(z_{2})=F_{\rm N}(z_{2})$.
 The respective electron probabilities $P(z_{1})$ and $P(z_{2})$
 at the end of the pulse ($t=5$~fs) are shown in Fig.~7(d).

 It is seen from the results presented in Figs.~7(c) and 7(d) 
 that the EER-induced ionization of atom~{\rm B} with the 
 entangled initial state proceeds similar to that with 
 the unentangled direct-product initial state described in Sec.~III above. 
 Specifically, [see Fig.~7(c)] the laser-driven
 electron $e_{1}$ (coordinate $z_{1}$) comes from atom~{\rm A} to atom~{\rm B}
 and pushes electron $e_{2}$ (coordinate $z_{2}$) out into the domain $z_{2}>z_{\rm B}$,
 where the EER-induced ionization of electron $e_{2}$ takes place in the positive
 direction of the $z_{2}$-axis.
 At the same time, electron $e_{1}$ which arrived at atom~{\rm B} does not affect 
 the probability $P(z_{1})$ at $z_{1}>50$~a.u..
 Similarly, [Fig.~7(d)], the laser-driven
 electron $e_{2}$ (coordinate $z_{2}$), coming from atom~{\rm A} to atom~{\rm B},
 pushes electron $e_{1}$ (coordinate $z_{1}$) into the domain $z_{1}>z_{\rm B}$,
 where the EER-induced ionization of electron $e_{1}$ takes place in the positive
 direction of the $z_{1}$-axis.
 Electron $e_{2}$ does not affect the probability $P(z_{2})$ 
 at $z_{2}>50$~a.u..
 In a realistic case, when both electrons are excited by the laser field [Figs.~7(a) and 7(b)],
 both  processes take place simultaneously.

 \begin{figure}[h]
 \includegraphics*[width=16pc]{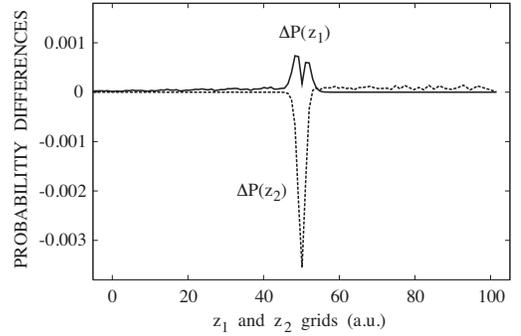}
 \caption{Excitation of H-H from the entangled initial state
 by the laser pulse with a narrow spatial envelope which affects only electron $e_{1}$
 in the domain of atom~{\rm A}:
 probability differences $\Delta P(z_{1})$ (solid line) and $\Delta P(z_{2})$ (dashed line)
 in the domain of atom~{\rm B}  
 at the end of the 5~fs laser pulse. 
 The changes of the electron probabilities $P(z_{1})$ and $P(z_{2})$ are measured 
 with respect to their initial values at $t=0$ [see Eq.~(\ref{E-DPs})].}
 \end{figure}

 To identify local maxima of electronic wave functions, 
 we next define the probability differences for electrons,
 $\Delta P(z_{1},t)$ and $\Delta P(z_{2},t)$, with respect to the initial electronic probabilities
 as follows:
 \begin{displaymath}
 \Delta P(z_{1},t) = P(z_{1},t) - P(z_{1},t=0),
 \end{displaymath}
 \begin{equation}
 \Delta P(z_{2},t) = P(z_{2},t) - P(z_{2},t=0).
  \label{E-DPs}
 \end{equation}
 The probability differences $\Delta P(z_{1})$ and $\Delta P(z_{2})$ plotted in Fig.~8 are
 calculated at the end of the 5~fs laser pulse with a narrow spatial envelope which excites 
 in the domain of atom~{\rm A} only electron $e_{1}$ and does not affect electron $e_{2}$.
 It can be concluded from Figs.~7(c) and 8 that
 the laser-driven electron $e_{1}$, coming to the domain $z_{1}>0$,
 is attracted and accelerated therein by proton $p_{\rm B}$ and forms at $t=5$~fs
 a local double-peak maximum at $z_{\rm B}=50$~a.u., 
 where electron $e_{2}$ already has its global maximum.
 Due to a very strong Coulombic repulsion of the two electrons in the vicinity of $z_{\rm B}=50$~a.u.,
 a sharp hole appears in the global maximum of $P(z_{2})$ at the expense of the increasing
 electron probability $P(z_{2})$ in the domain of $z_{2}>z_{\rm B}$, thus giving rise to
 the EER-induced ionization of electron $e_{2}$ in the positive direction of the $z_{2}$-axis.
 A similar process takes place when only electron $e_{2}$ is excited by the laser pulse 
 with a narrow spatial envelope, giving rise to the EER-induced ionization
 of electron $e_{1}$ in the positive direction of the $z_{1}$-axis.
 In a realistic case, when both electrons are excited by the laser field in the domain of atom~{\rm A}, 
 both described above processes take place simultaneously.

 We conclude from the results presented in Figs.~7 and 8 that the physical reasons of the energy
 transfer from atom~{\rm A} to atom~{\rm B} and the `sequential' ionization of atom~{\rm B} 
 in the case of the entangled initial state and a laser pulse with a narrow spatial envelope,
 which excites only electrons of atom~{\rm A}, 
 are similar to those underlying the case of the direct-product initial state:
 the Coulombic attraction of the laser-driven electrons by proton $p_{\rm B}$,
 and the short-range Coulombic repulsion of the two electrons in the vicinity of proton $p_{\rm B}$,
 where their wave functions strongly overlap.

 \section{Conclusion} 

 In the present work we have studied numerically the non-Born-Oppenheimer 
 quantum dynamics of two distant H atoms
 (being referred to as atoms {\rm A} and {\rm B})
 with an arbitrary large initial internuclear separation of $R=100$~a.u. (5.29~nm),
 which have been excited by spatially shaped laser pulses. 
 We have found
 an efficient energy transfer from one H atom to the other 
 and a `sequential' ionization of the latter, induced by a short-range EER.
 The short-range EER, taking place e.g. in the vicinity of proton $p_{\rm B}$ of atom~{\rm B},
 occurs due to the preceding long-range LIET from atom~{\rm A}
 to atom~{\rm B} enhanced by the Coulombic attraction and acceleration 
 of the laser-driven electrons by proton $p_{\rm B}$.

 Both unentangled direct-product atomic states and entangled molecular states 
 have been used as the initial states of the H-H system in our numerical simulations.
 The first case, unentangled atomic states, is more likely to arise in an experiment:
 at a gas pressure of 1~atm., for example, the interatomic distance is about 100~a.u..
 The second case, entangled molecular states, requires an additional step:
 dissociation of the H$_{2}$ molecule, or generation of entanglement between
 two individual atoms.
 The field strengths of the spatially shaped laser pulses were chosen
 such as not to induce a strong ionization and the respective decrease of 
 the overall norm of the wave packet of H-H. 
 In stronger fields and with longer pulses both the energy transfer and the `sequential' ionization
 can be more efficient than those presented above. 
 The parameters of the laser pulses used in the present work
 (in particular the pulse duration, $t_{p}=5$~fs, and especially the laser carrier frequency, $\omega=1.0$~a.u.) 
 are not yet optimal, and more efforts will be required in order to find optimal spatially shaped laser fields
 suitable for the most efficient energy transfer and `sequential' ionization.

 We have shown that in the case of a narrow spatial envelope of the applied laser field, 
 when only electrons initially belonging to atom~{\rm A} are excited by the laser field,
 the physical mechanisms of the energy transfer from atom~{\rm A} to atom~{\rm B} 
 and the `sequential' ionization of atom~{\rm B} are as follows:
 (i) the Coulombic attraction of the laser-driven electrons by proton $p_{\rm B}$ of atom~{\rm B},
 resulting in the formation of narrow local maxima of the electronic wave functions
 in the vicinity of proton $p_{\rm B}$,
 and (ii) the short-range Coulomb repulsion of the two electrons in the vicinity of proton $p_{\rm B}$,
 where their wave functions strongly overlap.
 In a more general case of a wide spatial envelope of the laser field, for example a broad Gaussian,
 when both {\rm A} and {\rm B} atoms are excited by the field simultaneously,
 the same processes occur in the opposite direction as well:
 the energy is also transferred from atom~{\rm B} to atom~{\rm A}
 and the EER-induced `sequential' ionization also occurs in atom~{\rm A}.

 Furthermore, we have shown that long-range entanglement of the initially 
 unentangled direct-product state is established by the interaction with a spatially shaped laser pulse.
 Long-range entanglement attracts considerable interest these days.
 For example, entanglement transfer from dissociated molecules to photons
 has been explored in
 \cite{ShapiroM:03pra.EntangTransfer},
 where the concept of transferring the quantum state of two dissociated fragments
 sharing internal-translational entanglement to that of two photons and vice versa
 has been put forward.
 A special kind of an entangled state of one electron and two protons
 in a H$_{2}^{+}$ molecule adiabatically stretched to $R=20-100$~a.u. (1.06-5.29~nm) 
 has been studied recently in
 \cite{BandrShon:pra.VPA.Entang}.
 A quantum-gate mechanism based on electron spins in coupled semiconductor quantum dots,
 separated by 40~nm, has been considered in
 \cite{BurkardDiVi:99prb.SpinEntang}
 as an important application of long-range entanglement.
 Such gates provide a source of spin entanglement and can be used for quantum computers.
 Experimental generation of entanglement between two individual $^{87}$Rb atoms 
 held in two optical tweezers, separated by 4~$\mu$m, has been reported in
 \cite{Wilk:2010prl.EntangExp}.
 Finally we mention that quantum entanglement among extended biomolecules has been explored in
 \cite{IshizakiFleming:2010NewJPhys.BiomolEntangTheor} 
 (see also references therein).
 The results obtained in 
 \cite{IshizakiFleming:2010NewJPhys.BiomolEntangTheor} 
 demonstrate that there exists robust quantum entanglement among 
 chlorophyll molecules under physiological conditions for the case
 of a single elementary excitation.

 The main characteristic features of the entangled molecular initial states in our H-H model, 
 as compared to the unentangled atomic direct-product ones,
 are as follows: (i) the immediate response of atom~{\rm B} to the laser excitation of atom~{\rm A}, 
 and (ii) the existence of fast out-of-phase oscillations 
 of the `atomic' energies $E_{\rm A}(t)$ and $E_{\rm B}(t)$ after the end of the applied laser pulse.
 These features should occur in a long-range quantum communication among distant
 quantum systems for the following reason. 
 The time-dependent `atomic' energies $E_{\rm A}(t)$ and $E_{\rm B}(t)$ 
 represent the information encoded in the laser pulse. 
 Therefore, if only one atom~{\rm A} of the entangled {\rm A}-{\rm B} pair 
 is excited by the laser pulse, the encoded information reaches atom~{\rm B} (receiver) 
 instantaneously and, moreover, can be decoded both during and after 
 the end of the pulse acting on atom~{\rm A}.

 We have found that in the case of a narrow spatial envelope of an applied laser pulse,
 which excites only electrons belonging to atom~{\rm A},
 the `sequential' EER-induced ionization of atom~{\rm B} is more efficient than the laser-induced
 ionization of atom~{\rm A} [see Figs.~2(f) and 6(e)].
 These results suggest important consequences for more complex quantum systems,
 e.g. those composed of three and more distant sub-systems, such as $\{{\rm A}_{n}\}, n=1,2,\ldots ,N$. 
 If we excite by a laser pulse with a narrow spatial envelope only sub-system~${\rm A}_{1}$, for example, 
 and if ionization probabilities obey ${\rm I}_{{\rm A}_{n}}<{\rm I}_{{\rm A}_{n+1}}$, 
 we can stimulate an efficient chain 
 of `sequential' ionization events with ever increasing probability in the total $\{{\rm A}_{n}\}$ system.
 The same may also apply to the energy transfer in the $\{{\rm A}_{n}\}$ system.
 To achieve such collective energy transfer and `sequential' ionization, 
 a narrow spatial shaping of the laser field as reported in
 \cite{Havermeier:10prl.ICD}
 needs further exploration.
 To achieve the condition
 ${\rm I}_{{\rm A}_{n}}<{\rm I}_{{\rm A}_{n+1}}$ in the $\{{\rm A}_{n}\}$ systems
 with a Gaussian spatial envelope of Eq.(\ref{E-G}),
 we can always choose $z_{0}<0$ such that
 the effective-field amplitudes obey
 ${\cal E}_{0}^{{\rm A}_{n}}>{\cal E}_{0}^{{\rm A}_{n+1}}$. 
 In this case we can anticipate that the condition ${\rm I}_{{\rm A}_{n}}<{\rm I}_{{\rm A}_{n+1}}$ 
 is fulfilled at the edge of the Gaussian envelope.

 Finally, at a shorter pulse duration and/or at a smaller laser carrier frequency,
 such that the number of optical cycles per pulse is $N_{c}<15$ (see Ref.
 \cite{Bandrauk:2002.PRA.HatomNc15}),
 the role of CEP-effects as explored previously in LIET
 \cite{Bandr:07prl.HpH2pBO}
 offers new avenues for energy and information transfer at long-range distances.
 
 \acknowledgments
 This work has been financially supported by the Deutsche Forschungsgemeinschaft 
 through the Sfb 652 (G.K.P., O.K.) which is gratefully acknowledged, 
 and A.D.B. thanks the Humboldt Foundation for the financial support through
 a Humboldt Research Award.


\begin{thebibliography}{99}

\bibitem{Cederbaum:97prl.ICD}
L.S. Cederbaum, J.Zobeley, and F. Tarantelli,
\newblock Phys. Rev. Lett. {\bf 79}, 4778 (1997).

\bibitem{Marburger:03prl.ICD}
S. Marburger, O. Kugeler, U. Hergenhahn, T. M\"{o}ller, 
\newblock Phys. Rev. Lett. {\bf 90}, 203401 (2003).

\bibitem{Jahnke:04prl.ICD}
T. Jahnke, A. Czasch, M.S. Sch\"{o}ffler, S. Sch\"{o}ssler, A. Knapp, M. Kasz, 
J. Titze, C. Wimmer, K. Kreidi, R.E. Grisenti, A. Staudte, O. Jagutzki, 
U. Hergenhahn, H. Schmidt-B\"{o}cking, R. D\"{o}rner,
\newblock Phys. Rev. Lett. {\bf 93}, 163401 (2004).

\bibitem{Morishita:06prl.ICD}
Y. Morishita, X.-J. Liu, N. Saito, T. Lischke, M. Kato, G. Prumper, 
M. Oura, H. Yamaoka, Y. Tamenori, I.H. Suzuki, K. Ueda,
\newblock Phys. Rev. Lett. {\bf 96}, 243402 (2006).

\bibitem{Lablanquie:07jcp.ICD}
P. Lablanquie, T. Aoto, Y. Hikosaka, Y. Morioka, F. Penent, K. Ito,
\newblock J. Chem. Phys. {\bf 127}, 154323 (2007).

\bibitem{Jahnke:10NaturePhys.ICD}
T. Jahnke, H. Sann, T. Havermeier, K. Kreidi, C. Stuck, M. Meckel, M. Sch\"{o}ffler,
N. Neumann, R. Wallauer, S. Voss, A. Czasch, O. Jagutzki, A. Malakzadeh,
F. Afaneh, T. Weber, H. Schmidt-B\"{o}cking, R. D\"{o}rner, 
\newblock Nature Phys. {\bf 6}, 74 (2010).

\bibitem{Havermeier:10prl.ICD}
T. Havermeier, T. Jahnke, K. Kreidi, R. Wallauer, S. Voss, M. Sch\"{o}ffler,
S. Sch\"{o}ssler, L. Foucar, N. Neumann, J. Titze, H. Sann, M. Kuhnel,
J. Voigtsberger, J.H. Morilla, W. Schollkopf, H. Schmidt-B\"{o}cking, 
R.E. Grisenti, R. D\"{o}rner,
\newblock Phys. Rev. Lett. {\bf 104}, 133401 (2010).

\bibitem{Sisourat:10NatPhys.ICD}
N. Sisourat, N.V. Kryzhevoi, P. Koloren\v{c}, S. Scheit, T. Jahnke, L.S. Cederbaum,
\newblock Nature Phys. {\bf 6}, 508 (2010).

\bibitem{Bandrauk:2002.PRA.HatomNc15}
S. Chelkowski and A.D. Bandrauk,
\newblock Phys. Rev. A {\bf 65}, 061802 (2002).

\bibitem{Bandr:08prl.H2d3}
A.D. Bandrauk, S. Chelkowski, S. Kawai, and H. Lu,
\newblock Phys. Rev. Lett. {\bf 101}, 153901 (2008).

\bibitem{Bandr:07prl.HpH2pBO}
A.D. Bandrauk, S.Barmaki, and G.L. Kamta,
\newblock Phys. Rev. Lett. {\bf 98}, 013001 (2007).

\bibitem{MayKuehn:2010book}
V. May, O. K\"{u}hn, 
{\it Charge and Energy Transfer Dynamics in Molecular System, 
2nd revised and enlarged edition},
\newblock (Wiley-VCH, Weinheim, 2004);
A. Nitzan,
{\it Chemical Dynamics in Condensed Phases}, 
\newblock (Oxford, University Press, 2006).

\bibitem{BandrShon:pra.VPA.Entang}
A.D. Bandrauk and N.H. Shon,
\newblock Phys. Rev. A {\bf 66}, 013401(R) (2002);
\newblock Phys. Rev. A {\bf 81}, 062101 (2010).

\bibitem{Doslic:06pra.VPA}
N. Do\v{s}li\'{c},
\newblock Phys. Rev. A {\bf 74}, 013402 (2006).

\bibitem{Paramon:05cpl.HH-HD.07cp.HH-HD-Muon}
G.K. Paramonov,
\newblock Chem. Phys. Lett. {\bf 411}, 350 (2005);
\newblock Chem. Phys. {\bf 338}, 329 (2007).

\bibitem{Rabitz:94.jcp.HF}
M. Kalu\v{z}a, J.T. Muckerman, P. Gross, and H.Rabitz,
\newblock J. Chem. Phys. {\bf 100}, 4211 (1994).

\bibitem{Eberly:04pra.Entang}
M.V. Fedorov, M.A. Efremov, A.E. Kazakov, K.W. Chan,
C.K. Law, and J.H. Eberly,
\newblock Phys. Rev. A {\bf 69}, 052117 (2004).

\bibitem{ShapiroM:03pra.EntangTransfer}
D. Petrosyan, G. Kurizki, and M. Shapiro,
\newblock Phys. Rev. A {\bf 67}, 012318 (2003).

\bibitem{BurkardDiVi:99prb.SpinEntang} 
G. Burkard, D. Loss, and D.P. DiVincenzo,
\newblock Phys. Rev. B {\bf 59}, 2070 (1999).

\bibitem{Wilk:2010prl.EntangExp}
T. Wilk, A. Ga\"{e}tan, C. Evellin, J. Wolters, Y. Miroshnychenko, 
P. Grangier, and A. Browaeys,
\newblock Phys. Rev. Lett. {\bf 104}, 010502 (2010).

\bibitem{IshizakiFleming:2010NewJPhys.BiomolEntangTheor}
A. Ishizaki and G.R. Fleming,
\newblock New J. Phys. {\bf 12}, 055004 (2010).

\bibitem{Kono:04cp.H2p.nearIR}
H. Kono, Y. Sato, N. Tanaka, T. Kato, K. Nakai, S. Koseki, and Y. Fujimura,
\newblock Chem. Phys. {\bf 304}, 203 (2004).

\bibitem{Corkum:93prl.Recollision}
P.B. Corkum,
\newblock Phys. Rev. Lett. {\bf 71}, 1994 (1993).

\bibitem{Bandrauk:09pra.Recollision}
K.-J. Yuan and A.D. Bandrauk,
\newblock Phys. Rev. A {\bf 80}, 053404 (2009).

\end{thebibliography}
\end{document}